 \documentclass[final,3p,times,twocolumn]{elsarticle}

 \usepackage{graphics}

 \usepackage{xcolor}
\usepackage{amssymb}
\usepackage{amsmath}
\usepackage{cases}
\usepackage{epsfig,subfigure}
\usepackage{graphicx}
\usepackage{epstopdf}

 \biboptions{comma,square,compress,numbers,sort}

\journal{Physics Letters B}

\begin{document}

\begin{frontmatter}



\title{First-order formalism and thick branes in mimetic gravity}


\author[a,b]{Qun-Ying Xie}
\ead{xieqy@lzu.edu.cn}
\author[c]{Qi-Ming Fu}
\ead{fuqiming@snut.edu.cn}
\author[b]{Tao-Tao Sui}
\ead{suitt14@lzu.edu.cn}
\author[b]{Li Zhao}
\ead{lizhao@lzu.edu.cn}
\author[d]{Yi Zhong \corref{cor4}}
\ead{zhongy@hnu.edu.cn}
  \cortext[cor4]{The corresponding author.}

\address[a]{School of Information Science and Engineering, Lanzhou University, Lanzhou
730000, P. R. China}
\address[b]{Lanzhou Center for Theoretical Physics $\&$ Research Center of Gravitation, Lanzhou University, Lanzhou 730000, P. R. China}
\address[c]{Institute of Physics, Shaanxi University of Technology, Hanzhong 723000, P. R. China}
\address[d]{Hunan Provincial Key Laboratory of High-Energy Scale Physics and Applications, Hunan University, Changsha 410082, P. R. China}

\begin{abstract}
In this paper, we investigate thick branes generated by a scalar field in mimetic gravity theory. By introducing two auxiliary super-potentials, we transform the second-order field equations of the system into a set of first-order equations. With this first-order formalism, several types of analytical thick brane solutions are obtained.
Then, tensor and scalar perturbations are analysed. We find that both kinds of perturbations are stable. The effective potentials for the tensor and scalar perturbations are dual to each other. The tensor zero mode can be localized on the brane while the scalar zero mode cannot. Thus, the four-dimensional Newtonian potential can be recovered on the brane.
\end{abstract}


\end{frontmatter}


\section{Introduction}

Modified gravity theories have obtained great development and performance
in the study of some unsolved problems in general relativity such as the
dark energy problem, the dark matter problem, the singularity problem, etc.
By isolating the conformal degree of freedom of general relativity,
Chamseddine and Mukhanov proposed a theory called mimetic gravity~\cite{Chamseddine:2013kea}.
This theory was studied from the view of variational principle~\cite{Golovnev:2014}.
It was shown that after introducing the scalar potential this conformal degree of freedom becomes dynamical and
can mimic cold dark matter~\cite{Chamseddine:2013kea,Chamseddine:2014vna}
or dark energy~\cite{Casalino:2018,Matsumoto:2016,Nojiri:2016ppu,Chamseddine:2018}
and can resolve the singularity problem~\cite{Brahma and Golovnev:2018} and the cosmic coincidence problem~\cite{Dutta:2018}.
The extension of mimetic gravity was used to investigate the inflationary solution~\cite{Mansoori:2010}.
In Ref. \cite{Izaurieta:2020}, the authors proposed mimetic Einstein-Cartan gravity and proved that
torsion is a non-propagating field in this mimetic gravity.
Mimetic gravity theory was also extended to Horava-like theory and applied to galactic rotation curves~\cite{Sebastiani:2016ras}.
It was also applied to other gravity theories such as $f(R)$ gravity \cite{Nozari:2019,Momeni:2015gka,Nojiri_2014,Odintsov:2015ocy,Odintsov:2015cwa,Odintsov:2015wwp,Oikonomou:2015lgy},
Horndeski gravity \cite{Arroja:2018,Arroja:2016,Cognola:2016gjy} and
Gauss-Bonnet gravity \cite{Capozziello:2013,Astashenok2015}.

On the other hand, in order to solve the gauge hierarchy problem and the cosmological
constant problem, Randall and Sundrum (RS) proposed that our four-dimensional world could be a brane embedded in five-dimensional space-time~\cite{Randall:1999ee}. With the warped extra dimension, it was further found that the size of extra dimension can be infinitely large without conflicting with Newtonian gravitational law~\cite{Randall:1999vf}. This charming idea has attracted substantial researches in particle physics, cosmology, gravity theory, and other related fields \cite{Davoudiasl:1999tf,Shiromizu:1999wj,Gherghetta:2000qt,Rizzo:2010zf,Yang:2012dd,Agashe:2014jca, Csaki:2000fc,DeWolfe:1999cp,Gremm:1999pj, Liu:2017gcn}.

Recently, one of the interesting works appeared in Ref.~\cite{Sadeghnezhad:2017hmr}, which
applied mimetic gravity theory to the thin RSII brane model~\cite{Randall:1999vf}. It was shown that the mimetic scalar field can mimic the dark sectors on the brane and explain the late time cosmic expansion in the favor of observational data, and it has the capability to explain initial time cosmological inflation ~\cite{Sadeghnezhad:2017hmr}.
Later, other related topics about thick branes in mimetic gravity were studied in Refs.~\cite{ZhongYi:2018,ZhongYi:2018JHEP, Bazeia:2020formalism,XiangQian:2020}.
Thick branes with the inner structure and the stability of the perturbations were first investigated in Ref.~\cite{ZhongYi:2018}. Besides, it is known that first-order formalism is a very powerful tool to obtain analytical brane solutions~\cite{Afonso:2006,Janssen:2008,Menezes:2014}. With this formalism, the second-order coupled field equations can be written as a set of first-order ones by introducing one or more auxiliary super-potentials. Very recently, Bazeia et al. used first-order formalism to find brane solutions in mimetic gravity~\cite{Bazeia:2020formalism,Bazeia:2020formalism2}.
In this paper, we would like to investigate thick branes in mimetic gravity with the help of first-order formalism by including two super-potentials. In order to show the systematicness and effectiveness of the first-order formalism on finding analytical brane solutions, we will utilize polynomial, period, and mixed super-potentials. We will also investigate the stabilities of the tensor and scalar perturbations as well as the relationship between the localizations of the tensor and scalar zero modes.

The paper is organized as follows. In Sec.~\ref{Sec2}, we give a review of the thick brane model
and reduce the second-order field equations to the first-order ones
by introducing two auxiliary super-potentials.
In Sec.~\ref{Sec3}, we obtain three types of analytical brane solutions
by considering different forms of super-potentials.
In Sec.~\ref{Sec4} and Sec.~\ref{Sec5}, we focus on
tensor and scalar perturbations, respectively.
Finally, the conclusion and discussion are given in Sec.~\ref{SecConclusion}.

\section{First-order formalism for thick brane models} \label{Sec2}

We consider thick branes generated by a scalar field in
five-dimensional mimetic gravity. The corresponding action is given by
    \begin{eqnarray}
        S\!=\!\int d^4x dy\sqrt{-g}\left( \frac{R}{2}
        + L_{\phi} \right),
        \label{action mgb1}
    \end{eqnarray}
where $R$ is the five-dimensional scalar curvature, the Lagrangian of the mimetic scalar field $\phi$ is
    \begin{eqnarray}
     L_{\phi}=\lambda\left[g^{MN}\partial_M \phi \partial_N \phi-U(\phi)\right]-V(\phi).        \label{Lagrangian}
    \end{eqnarray}
Here $\lambda$ represents the Lagrange multiplier, $U(\phi)$ and $V(\phi)$ are two potentials. In this paper, $x^M$ and $x^\mu$ denote respectively the five-dimensional bulk coordinates and the four-dimensional brane ones, where the indices $M,N,\cdots=0,1,2,3,5$ and $\mu,\nu,\cdots=0,1,2,3$.

The variations of the action (\ref{action mgb1}) with respect to
the background metric $g_{MN}$, the scalar field $\phi$, and the Lagrange multiplier $\lambda$
lead to the following field equations
    \begin{eqnarray}
        \label{var eom1}
        G_{MN}+2\lambda \partial_M \phi \partial_N \phi-L_{\phi}g_{MN}&=&0,    \\
        \label{var eom2}
        2\lambda\Box^{(5)}\phi+2\nabla_{M}\lambda\nabla^{M}\phi+\lambda \frac{\partial U}{\partial \phi}+\frac{\partial V}{\partial \phi}&=&0, \\
        \label{var eom3}
        g^{MN}\partial_M \phi \partial_N \phi-U(\phi)&=&0.
    \end{eqnarray}
Here the five-dimensional d'Alembert operator is defined as  $\Box^{(5)}=g^{MN}\nabla_{M}\nabla_{N}$.

The line-element ansatz for a flat brane is given by
    \begin{eqnarray}
        \label{brane metric1}
       ds^2 \!\!&=\!\!& e^{2A(y)}\eta_{\mu\nu}dx^{\mu}dx^{\nu}+dy^2 \\
            \!\!&=\!\!& a^2(y)\eta_{\mu\nu}dx^{\mu}dx^{\nu}+dy^2.
    \end{eqnarray}
Using the line-element (\ref{brane metric1}), and considering that the scalar field is
static and depends only on the extra-dimensional coordinate, we can get the following second-order nonlinear coupled
differential field equations
    \begin{eqnarray}
        \label{eom21}
        6A'^2 + \lambda\Big(U(\phi) + \phi'^2\Big) + V(\phi) &=& 0, \\
       \label{eom22}
        6A'^2 + 3A'' + \lambda\Big(U(\phi)- \phi'^2 \Big) + V(\phi) &=& 0, \\
       \label{eom23}
       \lambda \left(8A'\phi' + 2\phi'' + \frac{\partial U}{\partial \phi}\right)
        +2\lambda'\phi'+\frac{\partial V}{\partial \phi}&=&0,  \\
       \label{eom24}
       \phi'^2-U(\phi)&=&0.
    \end{eqnarray}
Here, the primes denote derivatives with respect to the extra-dimensional coordinate $y$.
It can be seen that it is not easy to analytically solve the above second-order field equations directly.
However, one can reduce them to first-order field equations by introducing two auxiliary super-potentials~\cite{Bazeia:2020formalism}
    \begin{eqnarray}
        Q = Q(\phi), ~~~~W = W(\phi), \label{QW}
    \end{eqnarray}
and providing the potential
    \begin{eqnarray}
        \label{V}
        V(\phi) =Q_{\phi} W_{\phi}-\frac{2}{3}W^2,
    \end{eqnarray}
where $Q_{\phi} = \frac{d Q }{d\phi}$ and $W_{\phi} = \frac{dW }{d\phi}$.
The resulting first-order field equations can be
written as
    \begin{eqnarray}
        A'        \!\!&=\!\!&  -\frac{1}{3}W(\phi),\label{W} \\
        \phi'        \!\!&=\!\!&  Q_{\phi}  ,\label{phi}\\
        U(\phi)       \!\!&=\!\!&  Q^2_{\phi},  \label{U} \\
\lambda(\phi) \!\!&=\!\!&  -\frac{1}{2} \frac{W_{\phi}} {Q_{\phi}}.  \label{lambda}
    \end{eqnarray}
These equations would be helpful to give analytical brane
solutions. One can see from (\ref{U}) that the scalar potential $U(\phi)$
is related with the super-potential $Q(\phi)$, which is a function of the scalar field $\phi$.
The other potential $V(\phi)$ is determined by the two super-potentials $Q$ and $W$ (see Eq. (\ref{V})).

Note that the first-order equations (\ref{W})-(\ref{U}) can be divided into two groups. One is Eq.~(\ref{W}) related with $A$ and $W$, the other is Eqs.~(\ref{phi}) and (\ref{U}) related with $\phi,\,Q,\,U$.
Therefore, these equations can be solved with different approaches by giving different combinations from ($A,\,W$) and $(\phi,\,Q,\,U)$. For example, we can choose $W(\phi)$ and $Q(\phi)$, or $W(\phi)$ and $\phi(y)$.

The energy density of the brane is given by
    \begin{eqnarray}
        \rho(y) = T_{MN}u^{M} u^{N},
    \end{eqnarray}
where $u^{M} = (u^0,0,0,0,0)$ is the velocity of a static observer. From the condition of the velocity $g_{MN}u^{M} u^{N} = -1$, we have $u^0=\text{e}^{-A}$. Thus, the energy density can be written as
    \begin{eqnarray}
        \rho(y) = \frac{1}{2}V(\phi(y)),  \label{rho}
    \end{eqnarray}
which shows that the potential $V(\phi)$ and the profile of the scalar $\phi(y)$ determine the distribution of the thick brane along the extra dimension.

On the other hand, in order to localize gravity on the brane, the
warp factor $\text{e}^{2A(y)}$ should tend to zero rapidly enough as $y\rightarrow\pm \infty$, such that the condition $\int e^{2A}dy < \infty$ is satisfied. This will be derived in section \ref{Sec4}. Usually, we consider the solutions with $e^{2A(y)}|_{y\rightarrow\pm\infty} \rightarrow e^{-2k|y|}$, which corresponds to the branes embedded in an AdS spacetime with a negative cosmological constant. It should be pointed out that the contribution of the cosmological constant has been included in the energy density (\ref{rho}) for this case. Therefore, the net energy density should be
    \begin{eqnarray}
        \rho(y) = \frac{1}{2}V(\phi(y)) -\Lambda.  \label{rho2}
    \end{eqnarray}

\section{The thick brane solutions} \label{Sec3}

Next, we will take focus on finding some specific analytical
solutions of the thick brane model by solving the first-order
equations (\ref{W})-(\ref{lambda}).
Our main motivation here is to show the systematicness and effectiveness of the first-order formalism,
which is also called the super-potential method~\cite{DeWolfe:1999cp,Gremm:1999pj,Bazeia2004}.

\subsection{Polynomial super-potentials}
\subsubsection{Solution I}

\begin{figure}[!htb]
\begin{center}
\subfigure[The warp factor]{
\includegraphics[width=3.7cm,height=2.7cm]{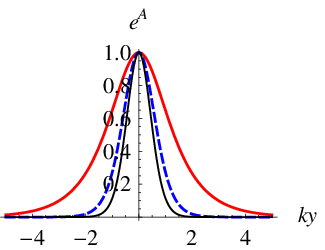}}
\subfigure[The  energy density]{
\includegraphics[width=3.7cm,height=2.7cm]{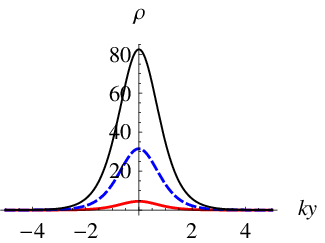}}
\end{center}
\caption{The shapes of the warp factor $a(y)$ and the energy density $\rho(y)$
 of the first brane model. The parameter $n$ is set as $n=1,\,3,\,5$ for the thick red, dashed
            blue, thin black lines, respectively.}\label{One}
\end{figure}

First, supposing that one of the super-potential $Q$ has the following polynomial form considered in Ref. \cite{Bazeia:2020formalism}:
    \begin{eqnarray}
        Q(\phi) =  k\left(v\phi - \frac{\phi^3}{3v}\right), \label{Model1_Q}
    \end{eqnarray}
and solving Eq.~(\ref{phi}), we can easily get the solution of the scalar field $\phi$
    \begin{eqnarray}
        \phi(y)= v\,\text{tanh}(ky).
    \end{eqnarray}
The solution of the potential $U$ can be read directly from Eq.~(\ref{U}) as
    \begin{eqnarray}
        U(\phi)  =  \frac{k^2 }{v^2} \left(\phi^2-v^2\right)^2. \label{U1}
    \end{eqnarray}
The other super-potential $W$ is chosen as
    \begin{eqnarray}
        W(\phi) = \frac{3kn}{v}\phi,
    \end{eqnarray}
where $n$ is a non-vanishing parameter. Then, from Eq.~(\ref{W}), we obtain a
simple solution for the warp factor
    \begin{eqnarray}
        A(y) = \ln\text{sech}^n(ky). \label{WarpFactor1}
    \end{eqnarray}
The other functions are given by
    \begin{eqnarray}
        \lambda(y) &=& -\frac{3n}{2v^2} \text{cosh}^2(ky), \\
        V(\phi)    &=& \frac{3nk^2}{v^2}\left[ v^2 -(2n+1) \phi^2 \right].
    \end{eqnarray}
The energy density of the brane including the cosmological constant is
    \begin{eqnarray}
        \rho^{(\Lambda)} = \frac{V}{2}= 3k^2n\Big(1 -(2n+1)\tanh^2(ky)\Big).
    \end{eqnarray}
From the solution (\ref{WarpFactor1}), we can calculate the cosmological constant and hence the net energy density
    \begin{eqnarray}\label{rho_y}
        \rho(y) = \rho^{(\Lambda)} -\Lambda = \frac{3}{2}k^2n(2n+1)\text{sech}^2(ky). \label{EnergyDensity1}
    \end{eqnarray}
The shapes of the warp factor $a(y)=\text{e}^{A(y)}$ and the energy density $\rho(y)$ are plotted in Fig.~\ref{One},
which shows that the parameter $n$ affects the warp factor and the energy density.
With the increase of $n$, the warp factor becomes narrower while the energy density becomes larger and narrower.
The maximum of the energy density is given by $\rho_{\text{max}}=\frac{3}{2}k^2 n(2n+1)$ for $n>0$ or $n<-1/2$.
It is obvious that the parameter $v$ does not affect the warp factor and the energy density, it only affects the amplitude of the scalar field $\phi$ and hence the localization of a bulk fermion $\Psi$ when one introduces the Yukawa coupling $\eta\bar{\Psi}\phi\Psi$~\cite{LiuYang2009}. In fact, $v$ is the vacuum expectation value of the scalar potential $U(\phi)$ given in (\ref{U1}).

\subsubsection{Solution II}

Next, we also consider the same polynomial super-potential $Q$
as in previous subsection
    \begin{eqnarray}
        Q(\phi)\!\!&=\!\!&  k\left(v\phi - \frac{\phi^3}{3v}\right).
    \end{eqnarray}
Therefore, we will get the same $\phi$ and $U(\phi)$ as solution I:
    \begin{eqnarray}
        \phi(y) &= & v\,\text{tanh}(ky),\\
        U(\phi) &= & \frac{k^2}{v^2} ({\phi^2}-{v^2})^2.
    \end{eqnarray}
But different from last model, we fix $W(\phi) = Q(\phi)$, which results in the constant solution of the Lagrange multiplier
    \begin{eqnarray}
        \lambda(y) = -\frac{1}{2} \frac{W_{\phi}} {Q_{\phi}} = -\frac{1}{2}.
    \end{eqnarray}
Then we get a different form of the warp factor via Eq. (\ref{W})
    \begin{eqnarray}
         A(y)  = \frac{v^2}{18}\text{sech}^2(ky)+\frac{2v^2}{9}\ln\text{sech}(ky).
    \end{eqnarray}
The scalar potential $V$ has the following $\phi^6$ form
    \begin{eqnarray}
        V(\phi) = \frac{k^2}{v^2}\left(\phi^2-v^2\right)^2 - \frac{2 k^2}{9  v^2} \left({\phi^2}-3v^2\right)^2 \phi^2.
    \end{eqnarray}
In this case, the energy density of the brane is
    \begin{eqnarray}
        \rho(y) &=& \frac{k^2v^2}{108}\Big( 27+10v^2 + 3(9+ 2v^2)\text{cosh}(2ky)\Big)\nonumber \\
                 &\times&     \text{sech}^6(ky).
    \end{eqnarray}
In this model, the vacuum expectation value $v$ of the scalar potential $U(\phi)$ has an explicit effect on the warp factor and the energy density. The shapes of the scalar field, the warp factor, and the energy density do not change with non-vanishing $v$.
Note that
    \begin{eqnarray}
        A(|y|\rightarrow\infty)
         &\rightarrow& \frac{2}{9} v^2 \left(e^{-2 k |y|}-k|y|\right)  \nonumber \\
         &\rightarrow& -\frac{2}{9} v^2 k|y|.
    \end{eqnarray}
Therefore, the five-dimensional spacetime is also asymptotic AdS.

\subsection{Period super-potential}
\subsubsection{Solution III}

Next, we try to construct another form of brane solution by giving
a period super-potential, such as
    \begin{eqnarray}
        W(\phi) = 3kn \sin^{\frac{1}{p}}\Big(\frac{\phi}{v}\Big).
    \end{eqnarray}
At the same time, the warp factor $A(y)$ is assumed to be
    \begin{eqnarray}
        A(y)=  \ln\text{sech}^n(ky). \label{WarpFactor3}
    \end{eqnarray}
Then the scalar field can be solved from Eq.~(\ref{W}) as
    \begin{eqnarray}
        \phi(y) =  v~\text{arcsin} [\tanh^p(ky)], \label{phiIII}
    \end{eqnarray}
which is a kink and a double kink for $p=1$ and $p=2n+1$ with positive integer $n$, respectively.
And from Eq.~(\ref{phi}) we know that the super-potential $Q$ is
    \begin{eqnarray}
        Q(\phi) = kp^2v^2
                 \left[ \mathcal{F}(p,\phi) + \mathcal{F}(-p,\phi) \right].
    \end{eqnarray}
where
    \begin{eqnarray}
        \mathcal{F}(p,\phi) &=& (-1)^{\frac{1}{2p}}  \sec^{\frac{1}{p}}\Big(\frac{\phi}{v}\Big) ~ \nonumber\\
        &\times&  {_2F_1}\left(\frac{1}{2p},\frac{1}{2p};1+\frac{1}{2p};\sec^2\Big(\frac{\phi}{v}\Big)\right).\nonumber
    \end{eqnarray}
Here ${_2F_1}$ is the hypergeometric function. The Lagrange multiplier and the two potentials can also be solved:
    \begin{eqnarray}
        \lambda(y)&=& \frac{ 3n\cot^2\Big(\frac{\phi}{v}\Big) \sin^{\frac{2}{p}}\Big(\frac{\phi}{v}\Big) }
                           { 2p^2v^2 \left[ \sin^{\frac{2}{p}}\Big(\frac{\phi}{v}\Big)-1
                                     \right ] },  \\
        V(\phi)&=& 3k^2n \left[1-(1+2n) \sin^{\frac{2}{p}}\Big(\frac{\phi}{v}\Big) \right],\\
        U(\phi)&=& k^2p^2v^2 \text{sech}^2 \Big(\frac{\phi}{v}\Big)\\
                &\times& \left[ \sin^{1+\frac{1}{p}} \Big(\frac{\phi}{v}\Big)
                               - \sin^{1-\frac{1}{p}}\Big(\frac{\phi}{v}\Big)
                         \right]^2.
    \end{eqnarray}
The energy density in this case is given by
    \begin{eqnarray}
        \rho(y) = \frac{3}{2}k^2n(2n+1)\text{sech}^2(ky), \label{EnergyDensity3}
    \end{eqnarray}
which is the same as Eq.~(\ref{EnergyDensity1}) for the first model. In fact, from the definition of the energy density $\rho=T_{MN}u^{M} u^{N}=(1/2)G_{MN}u^{M} u^{N}$, we know that the energy density will have the same configuration for the same warp factor. Here, the warp factors in this model and the first model have the same form, and hence so do the energy densities.

\subsubsection{Solution IV}

Similarly, for the  period super-potentials
    \begin{eqnarray}
         Q(\phi) = W(\phi)  =   k v^2 \sin\Big(\frac{\phi}{v}\Big),
    \end{eqnarray}
we can obtain the following solution
    \begin{eqnarray}
        A(y)   &=& \frac{v^2}{3}~\ln  \big[ \text{sech}(ky) \big],\\
        \phi(y)&=& v~\text{arctan} \big[\sinh(ky)\big],\\
        \lambda(y)&=& -\frac{1}{2}, \\
        V(\phi)&=& k^2 v^2 \left( \cos^2 \left(\frac{\phi}{v}\right)- \frac{2}{3} v^2 \sin^2 \left(\frac{\phi}{v}\right)\right),\\
        U(\phi)&=&  k^2 v^2 \cos^2 \frac{\phi}{v},\\
        \rho(y) &=& \frac{1}{6}k^2 v^2 (2v^2+3) \text{sech}^2(k y).
    \end{eqnarray}
%
%
\subsubsection{Solution V}

Next, we consider different period super-potentials  $W(\phi)$ and $Q(\phi)$:
    \begin{eqnarray}
         W(\phi)  &=&  3 k n \tan\Big(\frac{\phi}{v}\Big),\\
         Q(\phi)  &=& \frac{1}{2}kv^2 \sin \Big(\frac{2\phi}{v}\Big),
    \end{eqnarray}
The warp factor and the scalar field will have the following explicit forms
    \begin{eqnarray}
        A(y)   &=& \ln  \big[ \text{sech}^n(ky) \big],\\
        \phi(y) &=& v \arctan[\tanh(ky)].
    \end{eqnarray}
The other functions are given by
    \begin{eqnarray}
        \lambda(y)\!\!\!\!&=&\!\!\!\! -\frac{3n}{2v^2} \sec [ 2\arctan \tanh(ky) ] ,\nonumber\\
                           &\times& (1+ \tanh^2 (ky)) , \\
        V(\phi)\!\!\!\!&=&\!\!\!\! 3k^2n \sec^2 \left(\frac{\phi}{v}\right) \left[ (1+n) \cos \left(\frac{2\phi}{v}\right)-n  \right], \\
        U(\phi)\!\!\!\!&=&\!\!\!\! k^2 v^2 \cos^2 \Big(\frac{2\phi}{v}\Big).
    \end{eqnarray}

\subsection{Mixed super-potential}
\subsubsection{Solution VI}

Finally, we would like to generate the   brane model
by giving the pair of super-potentials $Q(\phi)$ and $W(\phi)$ as
the mixed of polynomial and period, for example
    \begin{eqnarray}
        Q(\phi)  =  W(\phi)  =   k v\Big[\phi + v \sin \Big(\frac{\phi}{v}\Big)\Big],
    \end{eqnarray}
which results in the constant Lagrange multiplier
    \begin{eqnarray}
        \lambda(y) = -\frac{1}{2}.
    \end{eqnarray}
Then governed by Eq.~(\ref{phi}), the scalar field  $\phi$ is determined as
    \begin{eqnarray}
        \phi(y)  =  2v \arctan(ky),
    \end{eqnarray}
from which one can see that the asymptotic behavior of $\phi$ is lim$_{y\to\pm\infty}$ $\phi(y)= \pi$.
The potentials $U$ and $V$ are
    \begin{eqnarray}
        U(\phi)\!\!\!\!\!&=\!\!\!\!\!& 4 k^2 v^2 \cos^4 \Big(\frac{\phi}{2v}\Big),\\
        V(\phi)\!\!\!\!\!&=\!\!\!\!\!&  k^2 v^2 \left[ 4\cos^4\Big(\frac{\phi}{2v}\Big) -\frac{2}{3}\Big(\phi + v\sin\Big(\frac{\phi}{v}\Big)\Big)^2\right].
    \end{eqnarray}
The warp factor and the energy density read as
    \begin{eqnarray}
        A(y) \!\!\!\!&=\!\!\!\!&  -\frac{2}{3} v^2  ky \arctan(ky),\\
      \rho(y) \!\!\!\!&=\!\!\!\!& \frac{1}{3}k^2v^2
            \Bigg[ \frac{6}{\left(k^2 y^2+1\right)^2} +\pi ^2 v^2
               \nonumber\\
                &-\!\!\!\!&  v^2 \Big(2 \arctan(k y)+\sin \left(2 \arctan(k y)\right)\Big)^2 \Bigg].
    \end{eqnarray}
The asymptotic behavior of $A(y)$ is $A(|y|\rightarrow \pm\infty\rightarrow -\pi  v^2 k|y|/3$ .

\section{Tensor perturbation} \label{Sec4}

\begin{figure}[!htb]
    \begin{center}
    \subfigure[~$$]{\label{figure tensor11}
        \includegraphics[width=3.7cm,height=2.7cm]{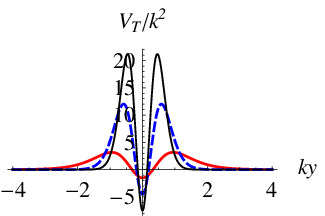}}
    \subfigure[~$$]{\label{figure tensor12}
        \includegraphics[width=3.7cm,height=2.7cm]{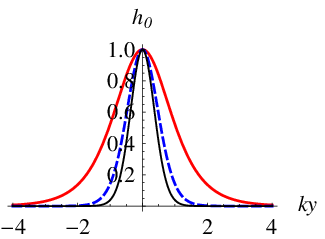}}
    \end{center}
    \caption{The effective potential $V_{\text{T}}$ and the non-normalized zero mode of the tensor perturbation
        for brane models I, III and V. The parameter is set as
             $n=1$ (red solid thick lines), $n=3$ (blue dashed lines), and $n=5$ (black solid thin lines).} \label{tensorThree}
    \end{figure}

\begin{figure}[!htb]
    \begin{center}
    \subfigure[~$$]{\label{figure tensor11}
        \includegraphics[width=3.7cm,height=2.7cm]{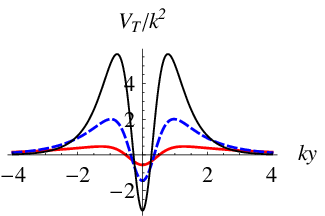}}
    \subfigure[~$$]{\label{figure tensor12}
        \includegraphics[width=3.7cm,height=2.7cm]{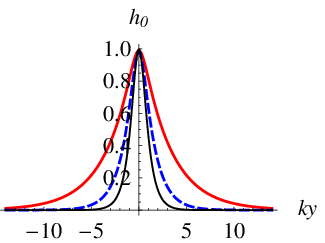}}
    \end{center}
    \caption{The effective potential $V_{\text{T}}$ and the non-normalized zero mode of the tensor perturbation
        for brane model II. The parameter is set as
             $v=1$ (red solid thick lines), $v=1.5$ (blue dashed lines), and $v=2$ (black solid thin lines).} \label{Three}
    \end{figure}

In this section, we consider the linear tensor fluctuation of the metric
around the background. Following the previous research works in
Refs. \cite{Csaki:2000fc,ZhongYi:2018,Bazeia:2020formalism}, we perform the following coordinate transformation
    \begin{eqnarray}
        \label{metric_transform}
        dz =e^{-A(y)}dy,
    \end{eqnarray}
to get a conformally flat metric
    \begin{eqnarray}
        \label{flat_metric}
       ds^2 \!\!&=\!\!& e^{2A(z)}(\eta_{\mu\nu}dx^{\mu}dx^{\nu}+dz^2) .
    \end{eqnarray}
To simplify the fluctuations of the metric
around the background, we only consider the
transverse and traceless part of the metric fluctuation, i.e.,
we consider the following tensor perturbation of the
metric:
\begin{eqnarray}\label{pertubation_metric}
 ds^2\!\!&=\!\!&\left(e^{2A(z)}\eta_{\mu \nu}
                +\hat{h}_{\mu \nu}(x,z)\right)dx^\mu dx^\nu
                +e^{2A(z)}dz^2\nonumber \\
    \!\!&=\!\!&e^{2A(z)}\left[\left(\eta_{\mu \nu}
                +h_{\mu \nu}(x,z)\right)dx^\mu dx^\nu
                +dz^2\right].
\end{eqnarray}
Here $ h_{\mu \nu}$ is the tensor perturbation of the metric, and it
satisfies the transverse and traceless conditions \cite{DeWolfe:1999cp}:
${h_\mu}^\mu=\partial^\nu h_{\mu \nu}=0$.
The non-vanishing  part of the perturbation of Einstein tensor in
Eq. (\ref{var eom1}) is the $\mu\nu$ components (since $h_{55}=0$ and $\delta G_{55}=0$) and it reads
    \begin{eqnarray}
    \label{perturbation Gmn}
        \delta G_{\mu\nu}\!\!\!\!&=\!\!\!\!&-\frac{1}{2}\Box^{(4)}h_{\mu\nu}+(6{A'}^2+3A'')e^{2A}h_{\mu\nu} \nonumber\\
                          &-& 2A' e^{2A}h_{\mu\nu}'-\frac{1}{2}e^{2A}h''_{\mu\nu} ,
    \end{eqnarray}
where the four-dimensional d'Alembertian is defined as $\Box^{(4)}\equiv\eta^{\mu\nu}\partial_{\mu}\partial_{\nu}$.
Using Eq.~(\ref{eom23}), we get the perturbation equation
    \begin{eqnarray}
        -\frac{1}{2}\Box^{(4)}h_{\mu\nu}
        - 2A' e^{2A}h_{\mu\nu}'-\frac{1}{2}e^{2A}h''_{\mu\nu}=0.
    \end{eqnarray}
Considering Eq.~(\ref{metric_transform}), we rewrite the above equation
under the coordinate $z$ as
    \begin{eqnarray}
        \Box^{(4)}h_{\mu\nu}
        + 3\dot{A}\dot{h}_{\mu\nu} + \ddot{h}_{\mu\nu}=0,\label{Eqs_of_h_mu_nu}
    \end{eqnarray}
where the dot represents the derivative with respect to the coordinate $z$.
By performing the following decomposition
  \begin{eqnarray}
  h_{\mu \nu}(x,z) =\varepsilon_{\mu\nu} e^{ikx}h(z)  e^{-\frac{3}{2}A},
    \label{decompsoseh_mu_nu}
  \end{eqnarray}
Eq. (\ref{Eqs_of_h_mu_nu}) leads to a kind of Schr\"{o}dinger equation
    \begin{eqnarray}
      \Big(-\partial_{z}^2+V_{\text{T}}(z)\Big) h(z)=m^2h(z),
    \label{KK_G_Schrodinger_Eqs}
    \end{eqnarray}
where $k^2=-m^2$ with $m$ the four-dimensional mass of a graviton KK mode,
and the effective potential given by
    \begin{eqnarray}\label{KK_G_potential_z}
    V_{\text{T}}(z)=\frac{3}{2} \ddot{A}(z)+\frac{9}{4} \dot{A}^2(z).
    \end{eqnarray}
The effective potential of the tensor perturbation under the physical coordinate $y$ is
  \begin{eqnarray}\label{KK_G_potential_y}
    V_{\text{T}}(z(y))=e^{2A(y)}\left(\frac{3}{2}A''(y)+\frac{15}{4}A'^2(y)\right).
  \end{eqnarray}
The zero mode of the tensor perturbation reads as
    \begin{eqnarray}
        h_0 (z)= e^{{3A(z)}/{2}}
            \left( c_1 +c_2 \int  e^{-3A(z)} dz \right) . \label{zeromode}
    \end{eqnarray}
This general form of the zero mode was first found in $f(R)$-brane model in Ref.~{\cite{CuiLiu2020}}.
One can see that the effective potentials and the zero mode of the tensor perturbation are only determined by
the warp factor $A(y)$. It is easy to show that the tensor zero mode (\ref{zeromode}) can be localized on the brane with the choice of $c_2=0$ for all brane solutions in this paper (with $n>0$ for solutions I, III and V and $v\neq0$ for solutions II, IV and VI). The figures of them
for solutions I, III and V are the same, and they
are similar for solutions II, IV and VI. So without loss of generality,
we plot two of them for the brane solutions I and II
in Figs. \ref{tensorThree} and \ref{Three}, respectively.
These potentials have a volcano-like shape. As the parameters $n$ and $v$ increase,
the potential wells in Figs. \ref{tensorThree} and \ref{Three} become narrower and deeper, respectively.

It is easy to verify that the zero modes for the above brane models are localized around the brane.
So the four-dimensional Newtonian potential can be realized on the brane.
There is no tensor tachyon mode, thus the brane is stable against the tensor perturbation.

\section{Scalar perturbation} \label{Sec5}

At last, we come to the scalar perturbation in this section.
The perturbed metric is given by
    \begin{eqnarray}
        \label{brane metric}
       ds^2 
           =e^{2A(z)}\left[(1+2\psi)\eta_{\mu\nu}dx^{\mu}dx^{\nu}+(1+2\Phi)dz^2\right].
    \end{eqnarray}
The scalar perturbation equations can be derived as 
        \begin{eqnarray}
      \!\!\!\!\!\! &e^{2A}& \!\!\!\! \left(\frac{1}{2}U_{\phi\phi}-U_{\phi}\frac{\ddot{\phi}}{\dot{\phi}^2}
                      +2 \frac{\dot{A}}{\dot{\phi}}U_{\phi}
                      -\frac{4}{3}e^{-2A}\lambda \dot{\phi}^2
                \right) \delta\phi  \nonumber \\
       \!\!\!\!\!\!&+&\!\!\!\! \left(\frac{1}{2}\frac{e^{2A}}{\dot{\phi}}U_{\phi}
                    +\frac{\ddot{\phi}}{\dot{\phi}}  - 2 \dot{A} \right) \dot{\delta \phi}
               -\ddot{\delta\phi}
               =0,
    \label{scalar master}
    \end{eqnarray}
and
    \begin{eqnarray}
        \Phi \!\!\!\!& =&\!\!\!\!  \frac{\dot{\delta \phi}}{\dot{\phi}}
                  -\frac{e^{2A}}{2 \dot{\phi}^2}U_{\phi}\delta\phi,  \label{Phi dt phi}\\
        \Phi \!\!\!\!&=&\!\!\!\!  - 2\psi  .  \label{ptb5}
    \end{eqnarray}
It can be seen that there is only one degree of freedom for the scalar perturbation.
Finally, by using the background equations (\ref{var eom1})-(\ref{var eom3}), we can replace the potential $U$ and $\lambda$ in Eq.~(\ref{scalar master}) with the functions $A$ and $\phi$ and obtain
the final form of the scalar perturbation $\delta \phi$:
    \begin{eqnarray}
      \!\!\!\!\!\!& \Bigg( & \!\!\!\!
             3\dot{A} \frac{  \ddot{\phi}}{ \dot{\phi}}
             -3 \ddot{A}
             -2\frac{\ddot{\phi}^2}{\dot{\phi}^2}
            +\frac{ \dddot{\phi} }{\dot{\phi}}
            \Bigg)\delta \phi   \nonumber\\
       \!\!\!\!\!\! &+& \!\!\!\!
            \left(2\frac{\ddot{\phi}}{\dot{\phi}}
                 -3\dot{A} \right)\dot{\delta \phi}
            - \ddot{\delta \phi}
            = 0 .  \label{scalar master2}
    \end{eqnarray}
Redefining
     \begin{eqnarray}
        \delta\phi(x^{\mu},z)=\overline{\delta\phi}(x^\mu) s(z) A^{-3/2}(z) \dot{\phi}(z) ,
    \end{eqnarray}
with the four-dimensional part of $\delta \phi$ satisfying $\Box^{(4)} \overline{\delta\phi}(x^\mu)=0$, 
we get the perturbation equation for the scalar degree of freedom $s(z)$ from Eq.~(\ref{scalar master2}):
     \begin{eqnarray}
        \label{scalar pertb}
        \left(-\partial^2_z +V_{\text{S}}(z)\right)s(z)=0,  \label{Eqsz}
    \end{eqnarray}
where the effective potential is given by
    \begin{eqnarray}
        V_{\text{S}}(z)= -\frac{3}{2} \ddot{A}(z)+\frac{9}{4} \dot{A}^2(z)      \label{Vsz}
    \end{eqnarray}
in the $z$ coordinate or
    \begin{eqnarray}
         V_{\text{S}}(z(y))= -e^{2A} \left( \frac{3}{2} A''(y) +\frac{3}{4}A'^2(y) \right)
            \label{Vsy}
    \end{eqnarray}
in the $y$ coordinate. Note that the corresponding effective potential of (\ref{Vsz}) in Ref.~\cite{ZhongYi:2018} is not right since there is an error in the matter equation (45) in that paper (the term $+2\lambda (\partial_z\phi)^2$ should be $-2\lambda (\partial_z\phi)^2$).

Comparing (\ref{Vsz}) with (\ref{KK_G_potential_z}),
one can see that the effective potential of the scalar perturbation $V_{\text{S}}$ is dual to
that of the tensor perturbation $V_{\text{T}}$: $V_{\text{S}}=V_{\text{T}}~(A\rightarrow -A)$, and Eqs.~(\ref{Eqsz}) and (\ref{KK_G_Schrodinger_Eqs}) can be rewritten as
    \begin{eqnarray}
         \mathcal{P}^\dag \mathcal{P} s(z) \!\!\!\! &=& \!\!\!\!0, \label{SchrodingerTensor}\\
         \mathcal{P} \mathcal{P}^\dag h(z) \!\!\!\! &=& \!\!\!\!0, \label{SchrodingerScalar}
    \end{eqnarray}
where $\mathcal{P}=\partial_z + \frac{3}{2}\dot{A}$. The above two equations ensure that both the scalar and tensor perturbations are stable. The zero mode can be obtained by replacing $A\rightarrow -A$ from the solution of the tensor zero mode (\ref{zeromode}):
    \begin{eqnarray}
        s (z)= e^{{-3A(z)}/{2}}
            \left( c_1 +c_2 \int  e^{3A(z)} dz \right) . \label{sz}
    \end{eqnarray}
This will lead to the conclusion that only one of the tensor and scalar zero modes can be localized on the brane.

\begin{figure}[!htb]
    \begin{center}
      \includegraphics[width=3.7cm,height=2.7cm]{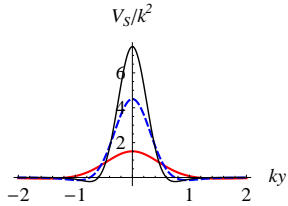}
    \end{center}
    \caption{The effective potential  $V_{\text{S}}(z(y))$ for solution I. The parameter is set as
              $n=1$ (red solid thick lines), $n=3$ (blue dashed lines), and $n=5$ (black solid thin lines).} \label{figVs_model1}
    \end{figure}

The effective potential (\ref{Vsy}) for solution I is given by
    \begin{eqnarray}
         V_{\text{S}} = \frac{3}{8} k^2 n \text{sech}^2(k y) \bigg(4+ n-n \cosh (2 k y)\bigg),    \label{Vsy1}
    \end{eqnarray}
which is shown in Fig.~\ref{figVs_model1}.
From this figure, it can be seen that, the value of the effective potential at $z=0$ will increase with the parameter $n$. This can be checked from the expression $V_{\text{S}}(0)= \frac{3 k^2 n}{2}$. The potential has
two very shallow wells, and approaches $0^{-}$ when $y \rightarrow \pm \infty $.

For the brane solution I in (\ref{WarpFactor1}) with $n>0$, it is easy to show that the zero mode (\ref{sz}) of the scalar perturbation cannot be localized on the brane. Thus, there is no additional fifth force coming from the scalar perturbation.
For other brane solutions, we have also the same conclusion.

\section{Conclusion} \label{SecConclusion}

In this work, we investigated the super-potential method
with which the second-order equations can be reduced to the first-order ones
for thick brane models in modified gravity with Lagrange multiplier.
The main step of this method is to introduce a pair of auxiliary super-potentials,
i.e., $W(\phi)$ and $Q(\phi)$. With these two super-potentials, the field equations are rewritten as Eqs.~(\ref{V})-(\ref{lambda}). Then we  try to use the method to find a series of
analytical brane solutions via some polynomial super-potentials, period super-potentials, and mixed super-potentials.
The warp factor has the same shape and the same asymptotic behavior at the boundary of the extra dimension for all those solutions, and all of these branes are embedded in five-dimensional AdS spacetime. The scalar field $\phi$ is a double kink for (\ref{phiIII}) with odd integer $p \geq 3$ or a single kink for other solutions. These shapes of the scalar field will affect the localization properties of fermions on the brane through the Yukawa coupling $\eta \bar{\Psi} \phi \Psi$ \cite{LiuYang2009}.

We also considered the tensor and scalar perturbations of the brane system. It was shown that both equations of motion of the perturbations can be transformed into Schr\"{o}dinger-like equations. Furthermore, these equations can be recast as the forms of (\ref{SchrodingerTensor}) and (\ref{SchrodingerScalar}), which show that both the perturbations are stable. The effective potential of the tensor perturbation is dual to that of the scalar perturbation. Therefore, only one of the tensor and scalar  zero modes can be localized on the brane. For all of our brane solutions, the tensor zero mode can be localized on the brane while the scalar zero mode cannot. Thus, the four-dimensional Newtonian potential can be recovered on the brane and there is no
additional fifth force contradicting with the experiments.

\section*{Acknowledgement}

This work was supported by the National Natural Science Foundation of China (Grant No. 11875151, No. 12047501, and No. 11705070). Yi Zhong was supported by the Fundamental Research Funds for the Central Universities (Grant No. 531107051196).


\end{document}